# Ultrasensitive Alzheimer's Disease Biomarker Detection with Nanopillar Photonic Crystal Biosensors

Guilherme S. Arruda[1], Katie Morris[2], Augusto Martins[2], Yue Wang[2], Sian Sloan-Dennison[3], Duncan Graham[3], Steven D. Quinn[2,4], Emiliano R. Martins[1], Thomas F. Krauss[2,4].

Filiation:
1 São Carlos School of Engineering, Department of Electrical and Computer Engineering, University of São Paulo, São Carlos – SP, 13566-590, Brazil
2 School of Physics, Engineering and Technology, University of York, York, YO10 5DD, United Kingdom.
3 Department of Pure and Applied Chemistry, University of Strathclyde, Technology and Innovation Centre, Glasgow G1 1RD, United Kingdom.
4 York Biomedical Research Institute, University of York, York, YO10 5DD, United Kingdom.

## Abstract:

The recent development of drugs able to mitigate neurodegenerative diseases has created an urgent need for biomarker tests that can be readily used by practitioners. Although biomarker detection directly in patients' blood is now possible, low-cost point-of-care tests remain a challenge, because relevant biomarkers, especially amyloid-$\beta$ (A$\beta$) peptides, are small, they occur at very low concentrations, and detecting a single marker is insufficient. Here, we demonstrate a photonic resonant sensor able to detect 0.2 pg/ml of $A\beta42$ and $A\beta40$ in 1% human blood serum, equivalent to 20 pg/ml in undiluted serum, which is the clinically required level. This high performance is achieved by combining gold nanoparticle amplification with a dielectric nanopillar photonic crystal structure in a dimer configuration, while also employing an immunoassay approach for high selectivity and specificity. The design combines high resonance Q-factor, amplitude and sensitivity, ideally suited for sensing. We also show the detection of A$\beta42$ and A$\beta40$ peptides in the same channel, which is highly relevant for assessing disease progress and opens a route towards multiplexing. Together with the handheld operation we have demonstrated previously, these photonic innovations make a major contribution to the ability to detect and monitor the progression of neurodegenerative diseases such as Alzheimer's.

## Introduction:

Recently, the detection of small blood-based biomarkers, including amyloid-$\beta$ (A$\beta$) peptides and phosphorylated tau protein variants, has become relevant for the early diagnosis of neurodegenerative diseases such as Alzheimer's Disease (AD) [1,2]. Early diagnosis enables pre-symptomatic treatment of AD, which potentially reduces progressive neurodegeneration and cognitive decline [3-5]. With the development of suitable antibodies, the detection of $A\beta$ can now also be achieved in blood [6,7], avoiding the need for highly invasive testing in cerebrospinal fluid or the use of costly Positron Emission Tomography (PET) scans. Nevertheless, detection is difficult because Aβ peptides are small (~3-4.5 kDa), clinically relevant concentrations are low (low pg/ml) and single biomarker detection is insufficient for clinical diagnosis [8].

Despite recent progress (mainly with electrochemical sensors [9-11]), no point-of care test for neurodegenerative diseases is available on the market yet. We also note that electrochemical sensors can achieve very low limits of detection, in some cases even reaching sub pg/mL concentrations in blood plasma [10,11]. However, real-world applications demand multiple criteria to be met simultaneously including scalability, robustness, user-friendliness, cost and multiplexing capabilities [12,13]. In this context, it is essential to detect multiple biomarkers in parallel and important to choose a modality that can easily offer multiplexing capability. Some of the most established highly performing laboratory techniques such as ELISA or chemoluminescence use optical techniques, so it is plausible to explore an optical modality for this purpose. Here, we introduce several innovations to the rich toolkit of nanophotonic resonant sensors and demonstrate, as a proof of principle, their capacity to provide a viable solution to the AD biomarker detection problem.

In the nanostructured photonic sensor space, both plasmonic and all-dielectric resonators have been studied extensively [14-17]. Such nanostructures allow for the label-free detection of specific molecules while also enabling surface imaging and the multiplexing of different biomarkers [18-20]. Furthermore, photonic resonant sensors are compatible with low-cost fabrication processes and can be implemented with minimal optoelectronic elements for the signal read-out [21,22], thus combining high-performance with low-cost. Detecting A$\beta$ peptides, however, to the best of our knowledge, remains a major challenge for this class of sensors, mainly due to their low molecular weight. For example, we recently demonstrated the detection of A$\beta$42 using a label-free interferometric approach [23]. However, despite achieving very low phase noise and detecting concentrations as low as 100 pg/ml, we were not able reach the very low pg/ml regime that is critical for diagnosing the early onset of AD.

An amplification strategy is therefore required. We note that gold nanoparticles offer an interesting option, as they are already widely used to amplify the response of lateral flow tests and have also been used with interferometry [24]. Previous attempts to use gold nanoparticles in conjunction with photonic crystals involved matching plasmonic and grating resonances [25], combining the high field confinement of the former with the high Quality-factor (Q-factor) of the latter. This approach is elegant from the physics perspective, but opens up issues in terms of application and translation, such as the increased loss of the plasmonic resonance and the need to fine-tune the size of the plasmonic nanoparticle to ensure resonance matching. Here, instead, we opt for a different physical concept, where the nanoparticles are only used as an index contrast enhancer, away from the plasmonic resonance. As we show in the following, this strategy contributes to achieving a much lower limit of detection than what was reported in [26], for example.

Regarding the photonic modality, we opt for the guided mode resonance (GMR) approach. This approach offers resonances with typical Q-factors around 200-1000, it can be implemented with a low-cost LED light source, read out with a simple CMOS camera and be realised in a handheld configuration [22]. This modality can therefore meet the high-performance low-cost paradigm that is so essential for realistic healthcare devices [12], especially as they have already shown low-pg/ml detection capability for protein biomarkers, even in complex biofluids [27], and also offer high multiplexing capabilities [12]. The LoD of such sensors is inversely proportional to the product of three resonance parameters [28]: Q-factor, the signal amplitude ($A$) and the sensitivity to refractive index change ($S$). By comparison, plasmonic nanostructures offer stronger field overlap with the analyte, and therefore higher $S$, but they tend to have a lower QAS product because their intrinsic absorption losses limit both amplitude and Q-factor [29]. The interesting question is therefore whether we can further improve the performance of GMR-based sensors by combining the high Q-factor and amplitude of the dielectric structures with the high polarisability of the gold nanoparticles. To this end, we explored a dielectric nanopillar geometry.

## Results

### Nanopillar Geometry - Design and Characterization

The GMR phenomenon is based on periodic structures, also known as photonic crystals. Photonic crystals using single nanopillar unit cells tend to have a lower Q-factor than their nanohole equivalent (see SI Section 1 for more details). Indeed, we confirmed this perception and observed Q-factors of only around Q≈50. (see Fig. S1e). We therefore used a dimer geometry, which adds another degree of freedom to control the properties of the structure. This strategy appears similar to breaking the symmetry of the unit cell of a structure which supports a bound state in the continuum (BIC) [30,31], yet we follow a very different approach; we start the design from first principles and aim to understand the properties of the structure via Fourier analysis. This approach highlights the fact that the coupling between radiating and waveguided modes is controlled via the gap distance between the dimer nanopillars (see SI Section 1 for more details). We can then easily tune the radiative (or design) Q-factor to satisfy the critical coupling condition required to obtain high resonance amplitudes that are so important for sensing [28]. The dimer configuration also circumvents the trade-off between Q-factor and sensitivity because its field profile is largely independent of the Q-factor (see Fig. S1h-i).

The resonant structure (Fig. 1a-d) consists of a rectangular array (periods $\mathrm{a}_x$ and $\mathrm{a}_y$ in the corresponding **x** and **y** directions, see Fig. 1a-b), of dimer cylindrical nanopillars (each with diameter $W$), patterned in a commercially available 100 nm thick amorphous silicon (aSi) on glass substrate (see Methods). The pillars are separated by a centre-to-centre distance $g_c$, as shown in Fig. 1b. The design of the nanopillar unit cell uses Fourier analysis and the understanding that the resonance is governed by two Fourier components, i.e. that the first order Fourier component controls the coupling between radiating and waveguide modes, whereas the second Fourier component controls the coupling between counterpropagating waveguide modes [32-35]. Typically, in nanohole gratings, a higher fill-factor (FF, i.e., the ratio between high to low index material in the unit cell) reduces the first-order component, which leads to a higher Q-factor, while in nanopillar gratings, it can only be used to a limited extent (see SI Section 1 for more details). The dimer structure opens another degree of freedom; by varying the distance between the two pillars, we can control the coupling between the radiating and waveguided modes – and hence the Q-factor – without changing the FF (see Fig. 1e, black solid line). The centre-to-centre distance of the pillars $g_c$, therefore controls the first Fourier component $\epsilon_{01}$, as follows (See SI Section 2 for a complete derivation):

$$\epsilon_{01} = 2(\varepsilon_{aSi} - \varepsilon_c)FF \frac{J_1\left(\frac{\pi W}{\mathrm{a}_y}\right)}{\frac{\pi W}{\mathrm{a}_y}} \cos\left(\frac{\pi g_c}{\mathrm{a}_y}\right) \quad (1)$$

where $\varepsilon_{aSi}$ and $\varepsilon_c$ represents, respectively, the permittivity of the aSi and cover material (water in our case) and $J_1$ is the first order Bessel function of the first kind. As $g_c$ approaches $\mathrm{a}_y/2$, $\epsilon_{01}$ initially decreases and eventually goes to zero in the limit that $g_c = \mathrm{a}_y/2$. Since the magnitude of $\epsilon_{01}$ directly relates to the coupling strength between radiating and the waveguide mode, the Q-factor diverges to infinity when $g_c = \mathrm{a}_y/2$ (see Fig. 1e, blue dashed line). At this point, the period of the structure is halved, and the mode can no longer couple to radiation, which effectively closes the cavity (Q-factor becomes infinite). Such diverging Q-factor behaviour is also a characteristic of BICs [36], but we emphasize that the dimer mode with infinite Q-factor is not a BIC; this mode does not belong to the continuum because it is a simple waveguide mode operating below the light line, that is, in the discrete part of the spectrum of eigenvalues (see SI Section 3 for detailed band diagrams).

As shown in Fig. 1c, the structure supports modes with high electric field confinement around the pillars, which are of particular interest for sensing applications as this field distribution leads to a particularly high sensitivity, as we discuss next. This mode can be excited by a perpendicularly incident **x**-polarized light and its resonance wavelength is directly proportional to the period along the **y** direction ($\mathrm{a}_y$). A Scanning Electron Microscope (SEM) micrograph of the fabricated array is shown in Fig. 1d.

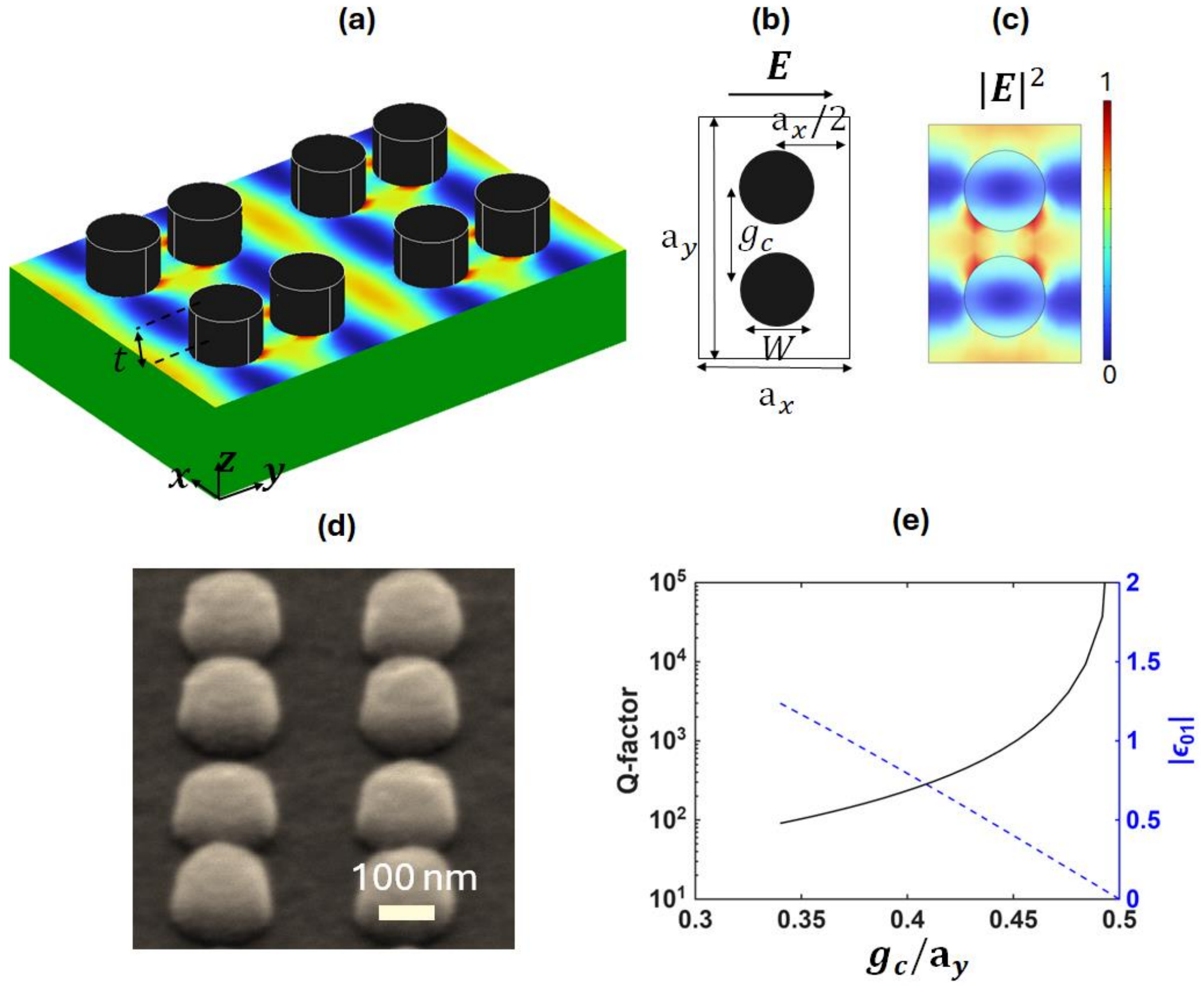

Figure 1: Amorphous silicon dimer nanopillar photonic crystal structure. (a) Schematic of the structure consisting of an array of aSi pillars (in black) of thickness $t = 100$ nm on top of a glass substrate (in green), together with the intensity distribution of the resonant mode. (b) Top view of the rectangular unit cell, with periods $\mathrm{a}_x = 320$ nm and $\mathrm{a}_y = 500$ nm in the respective $\boldsymbol{x}$ and $\boldsymbol{y}$ directions, indicating the centre-to-centre distance $g_c$ and the orientation of the incoming electric field $\boldsymbol{E}$. The diameter of the pillars is $W = 170$ nm. (c) Normalized intensity profile of the resonant mode for $g_c = 220$ nm with water as the cover. The mode distribution is calculated at the aSi-glass interface, i.e. at the bottom of the pillars, where the biomarkers are most likely to attach. (d) SEM micrograph of the fabricated structure consisting of a periodic array of aSi dimer pillars on a glass substrate. (e) Relationship between the mode's Q-factor (black solid line), unit cell Fourier component $\epsilon_{01}$ (blue dashed line) and distance between the pillars $g_c$ (horizontal axis). Note that the Q-factor diverges to infinity when $g_c$ approaches $\mathrm{a}_y/2$, which is the condition where $|\epsilon_{01}| = 0$.

## Figure of Merit for Sensing

In a typical nanohole photonic crystal [37], the Q-factor broadly scales with the fill-factor, which leads to a trade-off between the Q-factor and the sensitivity $S$: as the fill-factor increases, the mode becomes more confined, which in turn reduces $S$ (see Fig. S1b-c). The dimer configuration solves the trade-off problem because the field distribution is largely independent of $g_c$ (this effect is manifested as a constant effective index $n_{eff}$ value in the blue dashed curve in Fig. S1i – also see SI Section 4 for the dependence of the field energy on $g_c$). The fact that the electric field is highly confined outside the pillars then leads to a bulk sensitivity of 240 nm/RIU (see SI section 5 for experimental data), as shown in Fig. 2a (see Methods), which is considerably higher than regular photonic gratings or waveguide structures [21,38]. Moreover, the resonance Figure of Merit ($FOM$) scales not only with the Q-factor and $S$, but also with the amplitude of the resonance [28]. For convenience, we have slightly reformulated the expression derived in [28] and expressed it here in terms of the experimentally measured Q-factor ($Q$) rather than the ideal Q-factor as used before. As shown in SI Section 6, the $FOM$ then relates to the sensitivity, $S$, resonance amplitude, $A$, and Q-factor, $Q$, as:

$$FOM \sim SQ\sqrt{A} \tag{2}$$

Fig. 2 shows some examples and provides the data we used to estimate the $FOM$ of the dimer pillars, also highlighting the importance of the dimer spacing, here using $g_c$ = 190, 210 and 220 nm (see Fig.2 b-d, respectively). See SI Section 7 for the methods used for extracting the resonance Q-factor and A. We note that the Q-factor initially increases with $g_c$ at the expense of lower $A$ (Fig. 2b to Fig. 2c) and eventually reaches a maximum around 500-600 (Fig. 2c and Fig. 2d) due to the optical losses from the surface roughness of the structure. A measured Q-factor of 640 with an amplitude of 0.47 is, nonetheless, impressive for all-dielectric resonators in the near visible domain when compared to the literature [30,37,39]. A common perception is that, for the same sensitivity, a higher Q-factor implies higher $FOM$ and better LOD. However, this perception does not consider the signal amplitude, as the highest $SQ\sqrt{A}$ product does not always coincide with the highest $Q$-factor (see the SI Section 8 for an example). Thus, the dimer nano-pillars offer the opportunity to tune the resonance by varying $g_c$ and maximising the $SQ\sqrt{A}$ product; here we show that a value of $g_c = 210$ nm, see Fig. 2c, is optimum.

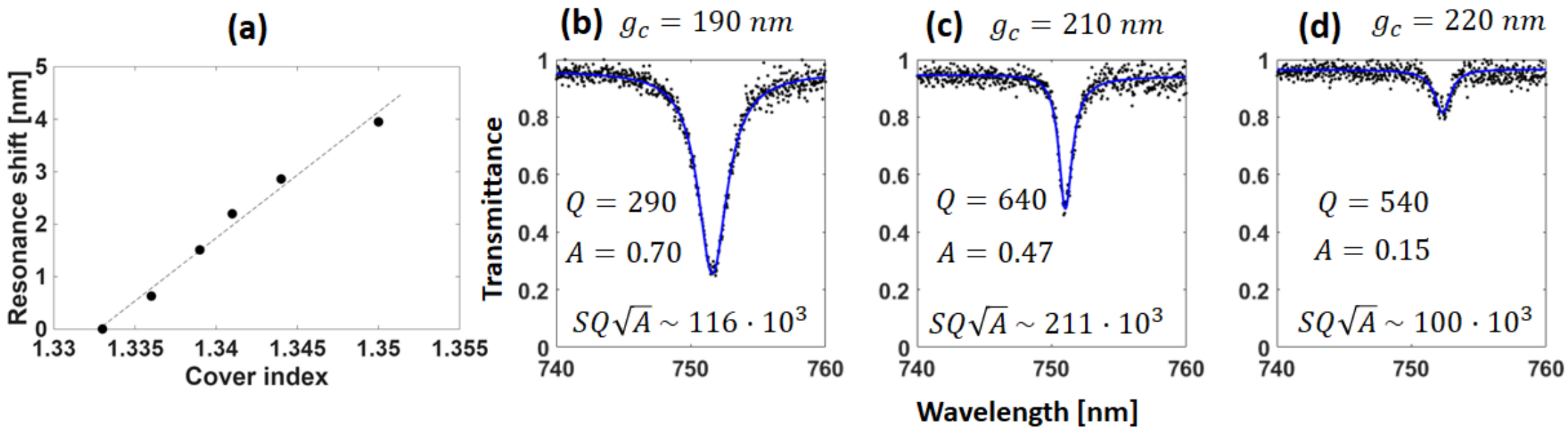

Figure 2: Optical characterization of the dimer pillar. (a) Resonance shifts of the mode highlighted in Fig. 1c for different values of cover indices, obtained by diluting ethanol in water. We determine a bulk sensitivity of 240 nm/RIU from the slope of the fitted curve. (b-d) Transmittance spectra measurements for three different $g_c$ values: 190, 210, and 220 nm. The black dots represent the measurement data while the blue solid curve is the Fano fitted curve used for extraction of the Q-factor and the amplitude $A$, which are displayed as inset values for each transmittance graph along their $SQ\sqrt{A}$ product (given in nm/RIU). We coated the sample in PMMA, which makes it easier to handle and measure the optical parameters, but it requires a slight detuning of the structural parameters to account for the difference in refractive index, such that the geometrical parameters of the dimer unit cell are: $a_x = 320$ nm, $a_y = 480$ nm and $W = 160$.

## Gold Nanoparticle Amplification

We achieve signal amplification by using gold nanoparticles (AuNPs) in a sandwich assay (Fig. 3a-b). Due to their high polarizability, the nanoparticles cause a much larger resonance shift than protein biomarkers alone; on the other hand, the price for this strong response is the intrinsically high loss of the metal, which may be detrimental for the high-Q GMR resonance. We also note that the 55 nm particles (see methods) support a localised surface plasmon resonance (LSPR) around 550 nm, which is far from our operating GMR resonance at 750 nm [40]. The key question is therefore whether the high polarizability of the AuNPs can be used to amplify high-Q resonances.

To answer this question, we first studied the detection of Immunoglobin-G (IgG, ~ 150 kDa, at a concentration of 50 $\mu g$/ml – see Methods for details) as a model system, with and without the addition of AuNPs. We used a chirped-bowtie configuration to translate the resonance spectral shift ($\Delta\lambda$) into a spatial shift ($\Delta x$). First, the grating is chirped by tapering $a_y$ along $\boldsymbol{x}$ between 496 nm and 504 nm over a distance of 500 $\mu$m, similar to [41]; then, another mirrored chirped grating is added next to the first one, thus obtaining a bowtie configuration, as illustrated in Fig 3c. When illuminated by normally incident monochromatic light (wavelength $\lambda_0 = 750\ nm,$ see methods), the resonance manifests itself as a bright bar (in red) spatially located in the region where the ratio $\lambda_0/a_y$ matches the mode's effective index ($n_{eff}$). The bars can then be detected by a digital camera (see methods), whereas the signal is measured as the difference in the relative distance of two resonant bars (one on either side of the bowtie grating) and tracked using a code that fits the resonance curves [21] (see SI Section 9 for details). We use a polydopamine-based surface functionalisation protocol with a blocking agent to minimise non-specific binding [42], see Methods. The resonance shift observed due to the addition of IgG was of a few hundreds of pm (black arrow in the inset of Fig. 3d), while the subsequent addition of AuNPs caused a shift of almost $\Delta\lambda = 5$ nm (blue arrow in Fig. 3d), which represents a significant amplification of one to two orders of magnitude. Furthermore, we note that the signal was acquired relatively quickly over the course of only a few tens of minutes.

Meanwhile, the quality of the resonance, even for such a relatively high protein concentration with a correspondingly high density of AuNPs interacting with the resonance, did not prohibitively degrade the measured signal, as seen in the resonance images taken during three different steps of the experiment: after the functionalisation of the surface with IgG antibodies (Fig. 3e), addition of the IgG antigen (Fig. 3f) and the AuNPs (Fig. 3g). In the chirped-bowtie grating configuration, the shift is measured as the difference in the relative distance of two similar resonances, both exposed to the biochemical reagents, as indicated in the red arrow in Fig. 3e-g [21]. Indeed, from the resonance curves (red) plotted in the graphs of Fig. 3e-g, the new $\sqrt{A}$, resonance square root of the amplitude, and Q-factor after the AuNPs addition are roughly 0.80x and 0.49x smaller, respectively (see SI Section 8 for extraction methods). As expected, the losses introduced by the binding of the AuNPs reduce the intensity and also broaden the signal. Nevertheless, from Fig. 3e-g, we observe a 27x amplification of the sensitivity provided by the AuNPs (an increase in relative spatial shift from $\Delta x_1 = 12\ \mu m$ to $\Delta x_1 = 320\ \mu m$, see Fig. 3f-g). Therefore, the new $FOM$, proportional to $SQ\sqrt{A}$, is at least 10x greater according to equation 2, meaning that it is indeed possible to use high-Q resonances together with AuNPs to enhance the performance of high-Q photonic resonant sensors.

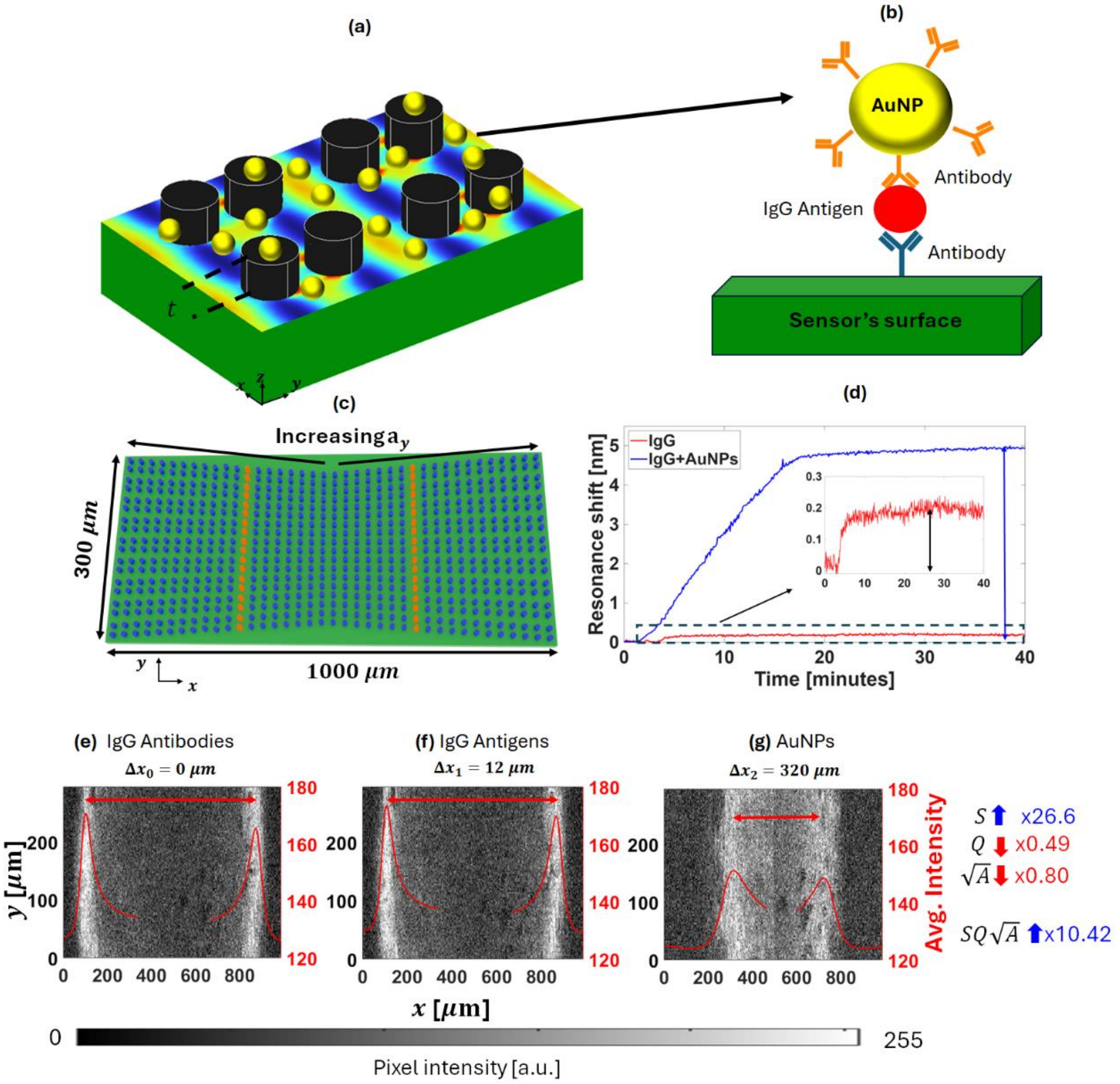

Figure 3: Sensitivity enhancement of AuNPs. (a) Schematic of the nanopillar dimer structure with the presence of the AuNPs (yellow circles) for sensitivity enhancement. (b) A schematic (not to scale) representing the AuNP assay: the IgG antigen (red circle) first binds to the IgG antibody on the sensor's surface (blue); in a second step, the antibody-functionalised (orange) AuNP (yellow circle) binds to the other terminus of the antigen. The polydopamine and blocking agent were omitted for simplicity. (c) Schematic of the chirped bowtie grating configuration where, upon excitation, the resonance manifests itself as two bright bars (highlighted in red). (d) Sensor response for IgG functionalisation and with nanoparticle amplification. A spectral shift of a few hundred pm (black arrow in zoomed-in inset graph) is observed for IgG binding alone, while a shift greater than 5 nm is measured with AuNP amplification (blue arrow). (e-g) Representative images of the sensor (raw data from the camera) at three different measuring stages: after the addition of the surface IgG antibodies (e), the IgG antigen (f) and functionalised AuNPs (g). Their respective fitted curves (the average pixel column intensity value plotted against its horizontal x position) are shown in red. The difference in the relative distances between the two bright resonant bars are $\Delta x_1 = 12\ \mu$m (f) and $\Delta x_2 = 320\ \mu$m (g). For the resonance extraction method, please see SI Section 9.

## Gold Nanoparticle Amplified Alzheimer's Disease Biomarker Detection

Next, we show that, as a proof-of-principle biosensor, the dimer nanopillar structure, in conjunction with AuNP amplification, is an extremely effective assay for the detection of Aβ peptides. We again use the chirped-bowtie configuration together with an aSi dimer pillar array (see SI Section 9). The assay protocol starts by functionalising the sensor surface with a coating of polydopamine, which alone forms a sticky layer that proteins can adhere to, without the engineering of specific binding sites being required. The downside of this strategy is that when antibodies attach to the polydopamine surface, they are randomly oriented. To overcome the randomness, we introduce protein G into the protocol. Protein G is a biotinylated antibody-binding protein which binds to the Fc region of antibodies and is widely used to optimize antibody orientation in laboratory based biomarker detection strategies such as ELISA [43,44]. By coating the polydopamine with a layer of neutravidin, the biotin side of protein G will selectively bind thereby orientating the antibody binding site away from the surface. Next, the anti-$A$β antibodies are introduced in-flow; they will be captured by protein G and orient correctly. Superblock® is then used as the blocking agent to occupy any remaining binding sites on the surface, thereby mitigating against non-specific binding. The $A\beta$ peptides are then flown over the sensor surface and bind to the correctly oriented antibody sites. See Methods for details.

The amplification strategy is designed to ensure that we can use the same AuNPs for both A$\beta$ peptides. To this end, we exploit the fact that the A$\beta$ peptides have a carboxyl (C-terminal) and an amino (N-terminal) end. The N-terminal is common to A$\beta$40 and A$\beta$42, while the C-terminal differs by the presence of two additional hydrophobic amino acids on the longer A$\beta$42 species. Therefore, the immobilised antibodies are chosen to selectively bind the C-terminal of the peptide, while the AuNPs are decorated with antibodies against the N-terminal. This way, the nanoparticles can bind to both types of peptides (Fig. 4a). See Methods for more details.

We start with the $A\beta$42 peptide and initially use laboratory buffer (PBS) to develop the assay. We run five concentrations of $A\beta$42 (20.0, 2.0, 0.8, 0.2 and 0.02 pg/ml) as well as pure PBS to control for non-specific binding. The AuNPs are then added and bind to the N-terminal of the amyloid. The response of each solution for 15 minutes after introducing the AuNPs is shown in Fig. 4b, with some representative images of the sensor and full-time plot shown in SI section 10. At 15 minutes we observe a shift of 0.2 nm for a concentration of 0.02 pg/ml (blue curve in Fig. 4b), which is 2x larger than the 0.1 nm non-specific shift in the control channel (black curve in Fig. 4b). Indeed, it is this shift of 0.1 nm that limits the LoD of the sensor, as it is greater than the noise level of the measurement, which has a standard deviation of $\sigma \sim 0.02$ nm (calculated from the sensor's response while PBS flushing). The curves present different shift rates (calculated from the linear regression of the resonance shift curves, see black dashed curves in Fig. 4a) for the first 15 minutes of the experiment, as shown in Fig. 4c (dots), which provides the important opportunity of using the dynamics rather than the saturation as readout. Additionally, we also fit the observed shift rates to a Langmuir curve (black line in Fig. 4c), a standard model for fitting single-site binding in biological systems.

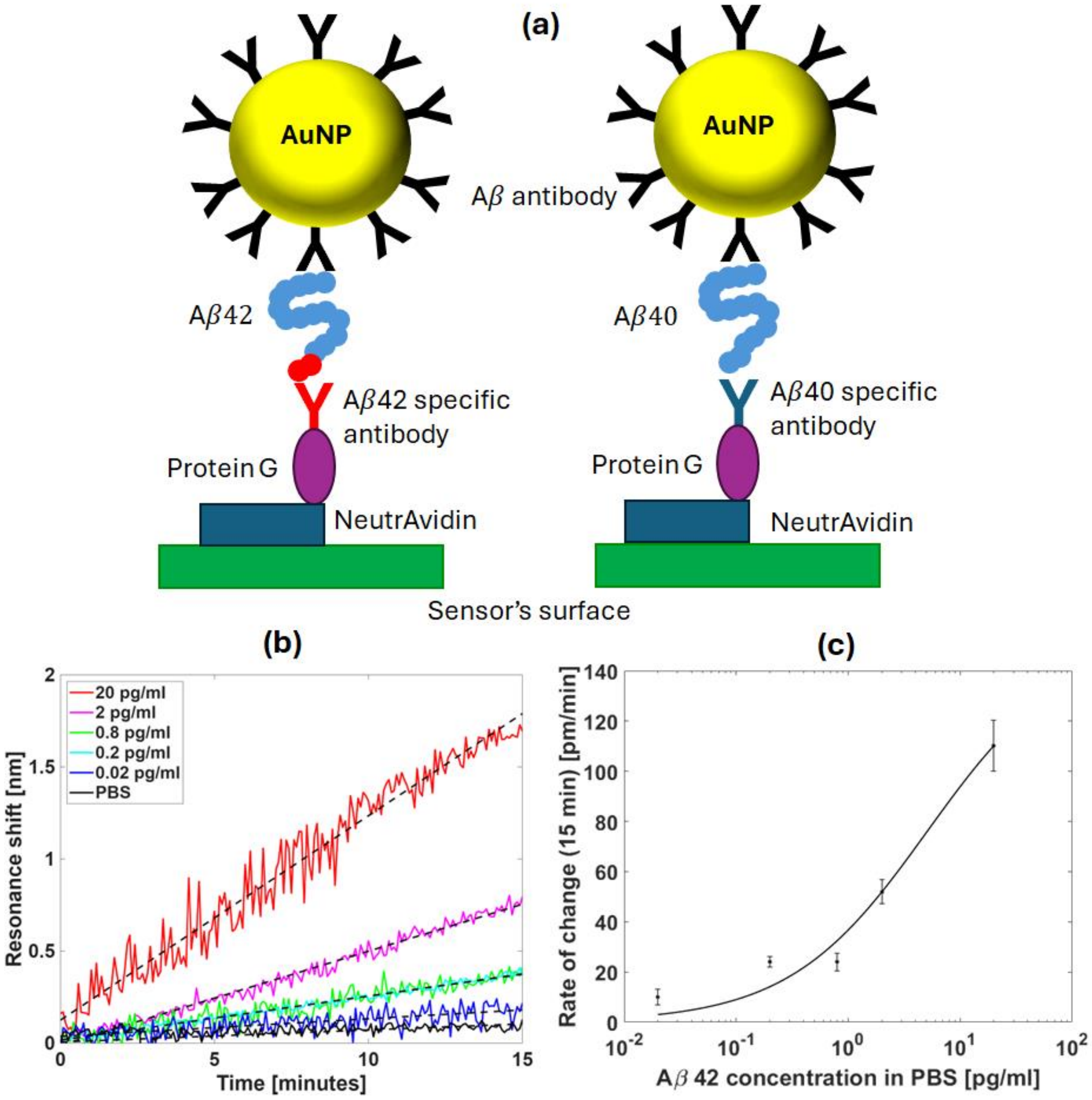

Figure 4: Detection of A$\beta$ peptides in a PBS solution. (a) Schematic (not to scale) of the surface chemistry used for capturing and detecting the analyte. Antibodies were immobilised and orientated using biotinylated protein G coupled to surface-bound neutravidin. The polydopamine and blocking agent were omitted for simplicity's sake. (b) Signal with the addition of the coated AuNPs for different A$\beta$42 concentrations in diluted in PBS: 20.0 (red), 2.0 (magenta), 0.8 (green), 0.2 (cyan) and 0.02 (blue) pg/ml; as well as pure PBS (black). The black dashed lines represent the fitted curve obtained via linear regression for each case. (c) Resonance shift rates at the initial 15 minutes of each experiment. The error bars represent the standard deviation from a moving window across the raw data divided by the time over which we calculate the gradient (15 minutes). The data was fitted to a Langmuir curve (black line). Each data point corresponds to a measurement performed on a freshly fabricated sensor. We observe a typical reproducibility of ±10% or better for nominally identical samples and experimental protocols.

## Blood-based Measurement and simultaneous detection of Aβ42 and Aβ40 peptides.

Finally, we demonstrate the detection of both Aβ42 and Aβ40 peptides simultaneously in the same channel, a requirement for the future quantification of these biomarkers and, consequently, the determination of their ratio. Since the long-term goal is to develop a finger-prick test, we dilute the serum 1:100 in PBS to provide sufficient liquid for microfluidic handling. The target concentration being 20 pg/ml amyloid in whole blood, as this is the clinically relevant level for AD [8], we need to detect 0.2 pg/ml in the dilution, and we do so for both Aβ42 and Aβ40. The results are shown in Fig. 5. We first test for each peptide separately, using spiked and unspiked dilution (see Methods). This comparison is necessary because of non-specific binding, as above, but also because of the nonzero concentration of A$\beta$ peptides that is present even in the blood of healthy individuals [8]. To quantify the background concentration, we performed a commercial ELISA test (see Methods). The resulting native concentration of A$\beta$42 in undiluted serum was 4 pg/mL and 60 pg/ml for A$\beta$40, which are of similar order as the 20 pg/ml equivalent concentration that we added. Both tests (Fig. 5a and Fig. 5b for A$\beta$42 and A$\beta$40, respectively) show a clear difference between the spiked and the unspiked cases, which highlights our ability to detect both the native background and any raised levels due to possible neurodegeneration. We note that the absolute resonance shifts we record for A$\beta$40 and A$\beta$42 are very different, which might be due to a difference in affinity of the corresponding antibodies. However, further investigation is required to confirm this assumption, especially as the binding affinity information is not provided by the manufacturer.

To detect both peptides at the same time, we run the Aβ42 and Aβ40 tests together in the same channel. In all cases, the relevant antibodies are introduced by spotting, which provides the ability to spatially separate different functionalisation chemistries and to run multiple tests in parallel (Fig. 5c), see Methods for more details. The experiment then proceeded as previously, except for the serum now containing both A$\beta$42 and A$\beta$40 at an additional spiked concentration of 20 pg/mL, on top of the native concentration of the peptides. The resonance shift curves are shown in Fig. 5d. The magnitudes of the shifts are comparable to those seen in Fig 5a for A$\beta$42 and Fig 5b for A$\beta$40 where the surface was functionalised in flow and the analytes only contained either A$\beta$42 or A$\beta$40 (compare the red and blue curves of Fig. 5a, 5b and 5d). This indicates that spotting the antibodies and detecting them simultaneously from the same analyte solution has little effect on the sensitivity of the measurement. Thus, our sensor has the potential to be used for simultaneous detection of multiple biomarkers, increasing the accuracy of AD diagnosis and potentially allowing for the differentiation of AD from other forms of dementia in the future.

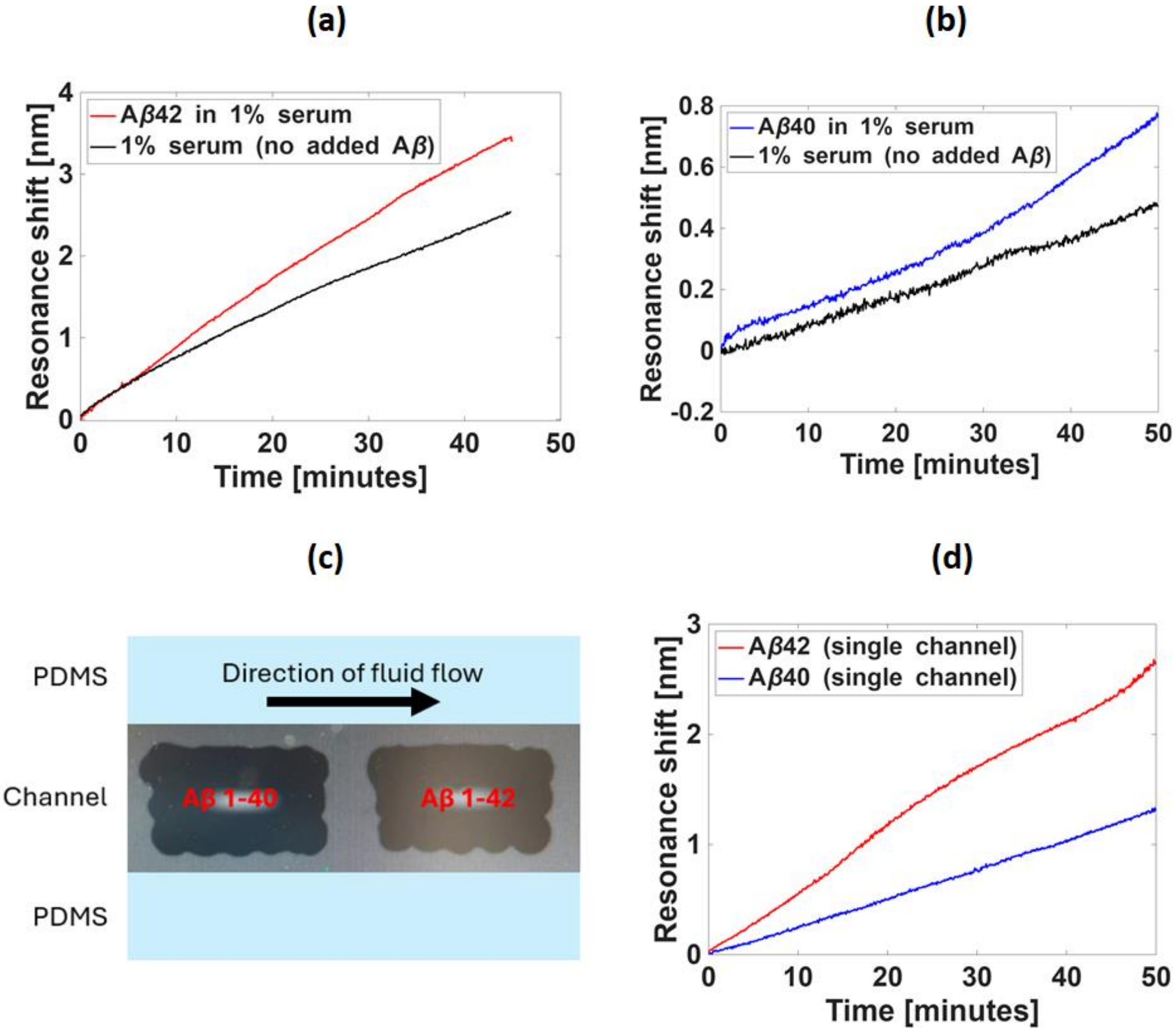

Figure 5: Detection of A$\beta$ in human serum. (a) Detection of 0.2 pg/mL A$\beta$42 spiked into diluted human serum. The sensor response is larger for the spiked (red) channel than for the unspiked channel (black). (b) Detection of 0.2 pg/mL A$\beta$-40 spiked into diluted human serum. Again, the sensor response is larger in the spiked (blue) channel than in the (black) unspiked channel. (c) Spotting region of the A$\beta$40 (left) and A$\beta$42

(right) antibodies over two sensing gratings (bright white rectangles) in a single fluidic channel. The blue borders define the microfluidic channels made of Poly(dimethylsiloxane) – PDMS. (d) Measured resonance shifts due to the flow of 0.2 pg/ml A$\beta$42 and A$\beta$40 spiked into the diluted human serum of dimer nanopillars functionalised with A$\beta$42 (red) and A$\beta$40 (blue) antibodies.

## Conclusions:

We have successfully demonstrated that gold nanoparticle-assisted photonic sensors can detect, simultaneously and within a single channel, two different A$\beta$ peptides, A$\beta$40 and A$\beta$42, at clinically relevant concentrations in human blood serum. In particular, we have been able to show the detection of 0.2 pg/ml of both peptides in 1% diluted serum, which is equivalent to 20 pg/ml in 100% serum and, thus, within clinically relevant levels for the early diagnosis of Alzheimer's disease; in this context, we note that we had to correct for their naturally occurring levels, which is of the order of a few pg/ml. Simultaneously detecting both A$\beta$42 and A$\beta$40 peptides is clinically extremely relevant as it opens up a viable route for quantifying their ratio and, thus, allowing monitoring of the progression of Alzheimer's disease, which would not be possible by monitoring the concentration of a single marker alone.

There are a number of novel insights that have contributed to this success. Firstly, we show that a lattice consisting of nanopillar dimers offers a sweet spot between the key parameters determining the performance of a photonic sensor, i.e. Q-factor, Sensitivity and Amplitude, leading to a very high Figure of Merit $FOM \sim QS\sqrt{A}$. Secondly, we demonstrate that gold nanoparticles can amplify the signal by one to two orders of magnitude when simply used as index contrast enhancers, despite the high Q-factor ($Q > 600$) of the guided mode resonance we use. Thirdly, we use the same functionalised nanoparticle to amplify two separate sensing regions (one for the detection of A$\beta$40 and one for A$\beta$42) by exploiting the difference between the C-terminus and the N-terminus of the A$\beta$ peptide that we are detecting, which allows us to simultaneously detect both A$\beta$42 and A$\beta$40 peptides within a single channel with a very simple assay. While no current biosensing platform fully meets all criteria for a point-of-care Alzheimer's biomarker testing, we note that the use of scalable materials (aSi on glass) and simplicity of the assay, together with the low-cost, handheld operation we have demonstrated previously [22], makes our modality a prime contender for translation into a near-patient test. Moreover, we have also demonstrated an all-passive microfluidic cartridge that draws plasma out of whole blood and across a GMR sensor to directly determine biomarkers in blood [45]. Nevertheless, more work is clearly required to demonstrate the full capability of the technology, yet we believe that our results establish a strong foundation for its translational development.

We suggest that a combination of all of these technologies is readily possible to make a portable, low-cost yet high-performance near-patient test that can be used to detect and monitor the progress of Alzheimer's Disease.

To provide even better diagnostic performance, we suggest extending the multiplexing capability further by including the detection of several phosphorylated variants of the tau protein into the assay, which will further improve the specificity of detection.

## Methods:

**Dimer array fabrication.** All sensors were fabricated using commercial wafers consisting of a 100 nm thick film of hydrogenated amorphous silicon (aSi) on a 500 $\mu$m glass substrate. The wafers were diced into 15 x 15 mm$^2$ pieces, which were then cleaned by sonication in acetone for 10 min, isopropanol for 5 min and then a dry $O_2$ plasma treatment for 5 minutes (100% power, 5 sccm $O_2$, *Henniker Plasma HPT-100*). An Alumina ($AlO_x$) hard mask was fabricated via a lift-off technique prior to the transferring of the dimer pattern to the aSi film. First, a 1:1 diluted in Anisole AR-P 6200.13 photoresist from *Allresist GmbH* was spin-coated on top of the substrate at 500 rpm for 5s, then spun at 3500 rpm for 45s, followed by a soft bake on a hot plate at 150 ºC for 2 min. The conductive polymer AR-PC 5090 (*Allresist GmbH*) was then spin-coated at 2000 rpm and baked on a hotplate at 90 °C for 2 min, which was necessary for charge dissipation during the EBL. The pattern was then defined by using electron-beam lithography (EBL, Raith GmbH Voyager, 50 kV) with a beam current of 900 pA and a dose of 160 $\mu$C/cm$^2$. After removing the AR-PC layer in deionized water at room temperature for 20 s, the exposed pattern was developed in Xylene for 55 s at room temperature and a quick rinse in isopropanol stopped the development. Next, a 30 nm thick $AlO_x$ layer was deposited using an electron-beam evaporator ($MBRAUN\ EVAP$). For the lift-off process, the substrate was soaked in 1165 resist remover (*Microposit*) on a hot plate at 70 ºC for 4 hours. Finally, the hard mask pattern was then transferred to the aSi film by plasma-based reactive ion etching (RIE) using a gas mixture of $SF_6$, $CHF_3$ and $O_2$ at a ratio of 20:12:13.5 for 55s at an acceleration voltage of 160 V and a chamber pressure of 0.1 mbar.

**Computational methods for field mode distribution and spectra calculations.** The Q-factor calculations in the SI and all the field distributions shown in this work were performed using the commercial software COMSOL Multiphysics "Eigenfrequency" study and the "Electromagnetic Waves, Frequency Domain" physics toolbox. Periodic boundary conditions were applied to the in-plane limits of the unit cell (plane XY of Fig. 1b), while scattering boundary conditions were used at the limits of the cover (water) and substrate (glass) materials. A rectangular simulation box of sizes $\Lambda_x$ x $\Lambda_y$ x 6 $\mu m$, with the periodic layer at the center level, was used. The transmittance spectra and the band diagrams shown in the SI 3 were obtained using an in-house implemented version of the Rigorous Coupled Wave Analysis (RCWA) method [34,46].

**Optical setup for the resonance parameters characterization.** The optical setup to characterize the resonance Q-factor and amplitude $A$ includes a collimated and coherent white light source (*LEUKOS SM-30*) and a high-resolution spectrometer (*Acton SpectraPro 2750* with an *Andor's Newton CCD*), in a transmission measurement configuration. Prior to the measurements, a 400 nm thick PMMA layer was spin-coated at 500 rpm for 5 s, and then 2000 rpm for 45 s, followed by a soft bake on a hot plate at 180 ºC for 5 min. The samples were mounted on a rotation stage, which allows for precise alignment. The transmission spectra were acquired by normalizing the spectrometer intensity response to the response of the same beam going through air.

**Chirped-bowtie grating characterization:** The optical setup to characterize the resonances of the chirped-bowtie gratings includes a tungsten-halogen lamp source (*ASBN-W High Power TH Light Source* from Special Products) together with a monochromator paired with a collimator unit (*Digitkrom*, Special Products) to generate the input beam and a digital camera (*Photometrics CoolSnap$^{TM}$ DYNO*, Digital Imaging Systems) to capture the sensor's image. The setup consists of an illumination and an imaging path, coupled by a beam-splitter. First, in the illumination path, the sample is excited by the collimated beam produced by a telescope consisting of a confocal (L1, with focal distance of $f_{L1}$ = 150 mm) and an objective (*RMS4X, Thorlabs* - with 4x magnification, numerical aperture NA = 0.1, focal distance of $f_{obj}$ = 45 mm and working distance WD = 18.5 mm). In the imaging path, the reflected beam is collected by the same objective and sent to a second lens (L2, $f_{L2}$= 180 mm) by means of the beam-splitter that acts as an inverted microscope imaging the grating on the camera sensor. A schematic of the setup can be found in Fig. S7d. A camera exposure time of 200 ms was used, and images were captured every five seconds (0.2 Hz).

**Bulk sensitivity experiments.** The measurements of the sensitivity of the sensor to changes in bulk refractive index were carried out using various dilutions of ethanol in water. The sensor was mounted onto the fluidic circuit consisting of a PDMS channel, an outlet tube connected to a syringe driver and an inlet tube. Solutions of ethanol in water from 10% to 60% were prepared in increments of 10% to cover a change in the refractive index from 1.33 to 1.35. The solutions were then introduced to the fluidics sequentially from lowest refractive index to highest refractive index and the shift of the resonance on the sensor was measured.

**Synthesis of the AuNPs.** The 55 nm (diameter) AuNPs were synthesized by adding sodium tetrachloroaurate (III) dihydrate (67.5 mg) to distilled water (500 mL) in a three-necked round bottom flask. The solution was heated, with constant stirring, until boiling. Once boiling, sodium citrate (60.5 mg) was added, and the solution heated for a further 15 minutes. It was then left to stir and cool for 12 hours. The resulting 0.4nM AuNPs solution was stored at room temperature until used.

**Functionalisation of nanoparticles.** To functionalise the 55 nm nanoparticles, 1000 $\mu$L of the AuNPs solution was added to 100 $\mu$L borate buffer, pH 9.0, along with 10 $\mu$L of an antibody which recognizes the N-terminal of Amyloid-β peptides (clone 6E8, 0.5 mg/mL, Genscript) for peptide sensing experiments (resulting in a final concentration of 4.5$\mu$g/ml), or 10 $\mu$L anti-IgG antibodies (anti-rabbit, 1 mg/mL, Sigma) for IgG detection (resulting in a final concentration of 9 $\mu$g/ml),. The nanoparticles were then agitated on a shaking plate for 60 minutes at 400 rpm in a glass vial at room temperature. 80 $\mu$L of 100 mg/mL BSA was added to the solution as a blocking agent and the nanoparticles were agitated for a further 30 minutes. The functionalised and blocked nanoparticles were then centrifuged at 4000 rpm for 20 minutes. The supernatant was discarded, and the pellet was resuspended in PBS for immediate use. 1 mL of nanoparticle solution was used per sensor, at a flow rate of 20 $\mu$L/min.

**Surface functionalisation for protein detection.** The sensor was first coated with a layer of polydopamine by submersion in a 2 mg/mL solution of dopamine HCl (Sigma) for 15 minutes. During polydopamine film formation the sensor was held vertically to avoid any debris in the solution settling on the film. The sensor was then washed with DI water and dried with nitrogen. Fluidics, consisting of a PDMS channel, an outlet tube connected to a syringe driver and an inlet tube were then assembled on top of the sensor. The fluidics were always operated with the direction of flow towards the syringe driver. For IgG detection, 1 mL of anti-IgG antibodies (anti-rabbit, Sigma) at a concentration of 50 $\mu$g/mL flowed across the surface. All flow rates were 75 $\mu$L/min unless otherwise specified. PBS was then washed through the fluidics for 10 minutes before 1 mL Superblock (Thermo) was used to block the surface. The channel was then washed again with PBS for 10 minutes before the introduction of IgG in PBS (see results). The channel was washed once more with PBS and the anti-IgG functionalised nanoparticles were used as the final amplification step at a flow rate of 20 $\mu$L/min.For Amyloid-β detection, 1 mL of NeutrAvidin (Thermo) at a concentration of 250 ug/mL was used as the initial layer, followed by 1 mL of biotinylated protein G (Thermo) at a concentration of 5 $\mu$g/mL to orientate the antibodies. 1 mL of anti-Aβ antibodies specific to either Aβ40 or Aβ42 (clone A40 or 25G13 respectively, Genscript) were then introduced to the channel at a concentration of 20 $\mu$g/mL. The surface was blocked using 1 mL superblock before the addition of the peptide-containing solutions in either PBS or diluted serum. The anti-A$\beta$ nanoparticles were then introduced to the channel as the final amplification step at a flow rate of 20 $\mu$L/min. The channel was washed with PBS between each step for 10 minutes at a flow rate of 75 $\mu$L/min.

**Spiking of human serum with $\boldsymbol{A\beta}$ peptides.** Amyloid-beta ($A\beta$) peptides were purchased as lyophilized powders from Anaspec. As the peptides have a propensity to aggregate, the peptides were first monomerized by dissolving the lyophilized powder in hexafluoroisopropanol (HFIP, Sigma) before aliquoting and desiccating into 20 $\mu$g aliquots [47,48]. When needed, an aliquot of the peptide was dissolved in 10 $\mu$L PBS containing 1% dimethyl sulfoxide (DMSO, Thermo) and the volume was made up to 1 mL with PBS. This solution was then diluted to the required concentration in PBS, along with human serum (Thermo) to a final concentration of 1%. Typically, this was 10 $\mu$L serum diluted to a total volume of 1 mL.

**Spotting of antibodies onto the sensor.** The sensor was coated with a layer of polydopamine by submersion in a 2 mg/mL solution of dopamine HCl (Sigma) for 15 minutes, as above. The sensor was then washed with DI water and dried with nitrogen before 100 $\mu$L of Neutravidin (1 mg/mL) was dropped onto the surface of the sensor and incubated for 10 minutes at room temperature. The sensor was then rinsed again with DI water and dried with nitrogen. 100 $\mu$L of protein G (20 ug/mL) was then dropped onto the surface of the sensor and incubated for 10 minutes at room temperature. After further washing in DI water and drying in nitrogen, the anti-Aβ antibodies were precisely spotted onto the sensors using a Scienion sciFLEXARRAYER S3 at a concentration of 300 $\mu$g/mL. Prior to spotting, the antibodies were degassed under a vacuum for 10 minutes. Once spotted, the antibodies were incubated on the surface for 1 hour at room temperature. The sensor was then rinsed in a large volume of PBS to remove unbound antibodies, before a final rinse in DI. The sensor was then dried with nitrogen before being assembled into the fluidics as described above. Superblock was flowed through the fluidics to block the surface prior to sensing. The remainder of the experiment was carried out as for the sensors that were functionalised in flow.

## Acknowledgements:

The authors G.S.A and E.R.M acknowledge financial support by the São Paulo Research Foundation – FAPESP (Grants #2020/15940-2, #2023/08797-7 and #2020/00619-4) and the Brazilian National Council for Scientific and Technological Development – CNPq (307602/2021-4). T.F.K, A.M, Y.W, S.S.D and D.G acknowledge funding from the Engineering and Physical Sciences Research Council – EPSRC (Grants EP/X037770/1 and EP/V047663/1). Prof. T. F. Krauss acknowledges support from the Wellcome Trust (221349/Z/20/Z). S. D. Q. thanks Alzheimer's Research UK (RF2019-A-001) for support. S. D. Q., T. F. K. and K. M. also acknowledge funding from the Alzheimer's Society (AS-PG-22-043) and the Alzheimer's Research UK Yorkshire Network.

## Data and code availability:

The authors declare that all the data and code supporting the findings of this study are available within the article, or upon request from the corresponding author.

## Competing interests statement:

The authors declare no competing interests.

## Additional information:

References [49,50] are cited only in the supplementary information (SI).